\documentclass{elsart}
\usepackage{psfig}
\usepackage{amstex}
\usepackage{amssymb}
\usepackage{cite}

\newlength{\hepwidth}
\begin{document}

\begin{frontmatter}

  \title{{\normalsize \hfill IFUP-TH 41/96\\
      \hfill hep-lat/9607049}\\[5ex]
    \bf Hybrid Monte Carlo and topological modes of full QCD}
  
  \author{B. All\'{e}s,}
  \author{G. Boyd\thanksref{corr},}
  \author{M. D'Elia,}
  \author{A. Di Giacomo}
  \author{\and E. Vicari}
  
  \address{Dipartimento di Fisica dell'Universit\`a and I.N.F.N.,
    I-56126 Pisa, Italy}

  \begin{abstract}
    We investigate the performance of the hybrid Monte Carlo algorithm, the
    standard algorithm used for lattice QCD simulations involving fermions, in
    updating non-trivial global topological structures.  We find that the
    hybrid Monte Carlo algorithm has serious problems decorrelating the global
    topological charge.
    This represents a warning which must be seriously considered when
    simulating full QCD by hybrid Monte Carlo.
  \end{abstract}
  
  \thanks[corr]{Corresponding author.  
    email:boyd@@ipifidpt.difi.unipi.it}
\end{frontmatter}

\newpage

\section{Introduction}
The present state of the art algorithm for lattice QCD with dynamical fermions,
used in all production runs, is the hybrid Monte Carlo~\cite{Weingarten,GT187,Kennedy}
(HMC) algorithm (for four staggered or two Wilson fermions), or the hybrid
molecular dynamics variant of this for fewer fermions.  In this letter we
investigate the effectiveness of the HMC algorithm, both with and without
fermions, in updating the topological sector. For the former we use four
flavours of staggered fermions. For the latter case we compare the HMC with
over-relaxed heat-bath results.

We find that decorrelating the topology requires a dramatic increase in
computer time as the quark mass decreases, over and above the increase naively
expected. For the smallest mass, the algorithm requires days to decorrelate the
topology, running on a powerful computer (25Gflop APE/QUADRICS).  This implies
that one cannot generate a set of configurations with the correct sampling of
the topological sector using brute force.

The correct sampling of topological modes\footnote{We regard the topological
  modes to be those which, in the continuum limit, determine the topological
  properties of the lattice.} is clearly crucial for quantities such as the
$\eta^{\prime}$ mass, which depend directly on the topology via the explicit
breaking of the $U_{\text{A}}(1)$ symmetry. It can also become relevant for
other quantities, for example the proton mass, at some level of accuracy.  Some
attempts to study the topology in the presence of fermions have been presented
in the literature~\cite{Bitar,fermcm,japan}.

\section{Topology}
This investigation began as part of a calculation of the spin content of the
proton~\cite{MTCspin}. This can be related to the matrix element of the
topological charge density operator $q(x)$ between on-shell proton states with
small momentum transfer. To this end we prepared a large set of configurations
using the HMC algorithm. We examined the longest auto-correlations in this set,
namely the auto-correlation time of the global topological charge $Q$. Any
reasonable definition of $Q$ is sufficient to determine the auto-correlations;
we used the field theoretic definition of $Q$, measured on
cooled~\cite{coolingmethod} configurations.

The global topological charge $Q$ is defined as
\begin{equation}
Q= \sum_x \frac{-1}{2^4\times 32 \pi^2}
\sum^{\pm 4}_{\mu\nu\rho\sigma=\pm 1}
\epsilon_{\mu\nu\rho\sigma} \text{Tr}
\left[ \Pi_{\mu\nu}(x)\Pi_{\rho\sigma}(x)\right].
\label{Q}
\end{equation}
where $\Pi_{\mu\nu}(x)$ is the plaquette in the $\mu\nu$ plane.  A related
quantity is the topological susceptibilty $\chi$ defined as
\begin{equation}
\chi = \langle Q^{2}\rangle / V_{4}
\label{chi}
\end{equation}
where $V_{4}$ is the lattice four-volume. The cooled configurations are
obtained from each configuration generated by the HMC by minimizing locally the
gauge part~\cite{fermcm} of the action.

Note that the conclusions drawn in this letter about the decorrelation of
global topological charge by the HMC algorithm are independent of the
definition of $Q$ used here.  It is sufficient to assume that the cooling
procedure is able to capture the most slowly updated modes associated with the
`topological' content of the lattice configurations, a perfectly reasonable
assumption. Possible systematic errors in this procedure will not change our
final conclusions.

As mentioned above, in order to extract physical results related to the
topological properties, a Monte Carlo simulation must provide a set of
configurations covering phase space. In particular, the global topological
charge must average to zero. So the algorithm must be effective in changing
topological modes. As we shall see, this efficiency appears unobtainable for
values of $\beta$ large enough to have scaling when the quark mass is taken as
small as currently feasible (eg., $m=0.01$). Note that this dramatic slowing
down is seen for quark masses still far from the chiral limit.

Although the existence of long auto-correlations has been
observed~\cite{Bitar,japan} in previous studies using the HMC algorithm,
results on many quantities, some associated with the topology, have been
reported.  We believe that critical slowing down in the HMC has been
under-estimated with respect to the topological modes.  As we shall show in the
following, this casts serious doubt on the possibility of performing any study
involving topological properties in the relevant region of $\beta$ and $m$.

\section{Results}
We have used the Wilson action for the pure gauge sector, and four flavours of
staggered fermions. Having four flavours allows the use of an exact algorithm,
the so-called $\Phi$ algorithm described in detail in reference~\cite{GT187}.
The molecular dynamics equations of motion are integrated in fictitious time
$\tau$, using the leapfrog integration method, and beginning with a half-step
in the gauge fields. The two relevant units for comparing the auto-correlations
from different simulations are then fictitious time in units of $\tau$, and
wall clock time.  The production runs were done using a $16^{3}\times 24$
lattice at a coupling $\beta=5.35$ for full QCD, and a $16^{3}\times 16$
lattice at $\beta=6.00$ for pure SU(3).

\begin{figure}[tb]
\begin{center}
\leavevmode
\psfig{figure=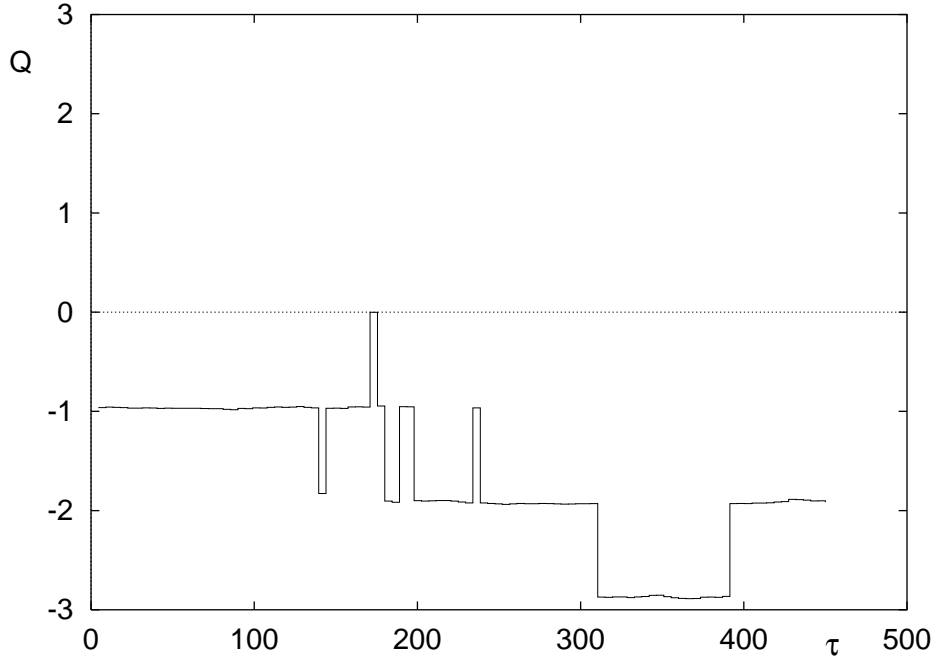}
\caption{
  Time history, in units of molecular dynamics time $\tau$, of the topological
  charge $Q$ for the HMC simulation at $\beta=5.35$ and $m=0.01$ on a
  $16^3\times 24 $ lattice.  }
\label{fig1}
\end{center}
\end{figure}

Let us begin with the results for quark mass $m=0.01$. This yields a pion to
rho mass ratio of $m_{\pi}/m_{\rho}\simeq 0.5$~\cite{MTC92}, so we
are still quite far from the physical value.  The lattice spacing is $a\simeq
0.14$fm (from $m_{\rho}$), which gives a lattice volume of
$V_{3}\simeq(2.2\text{fm})^3$.  The parameters used for the HMC simulation were
\begin{equation}\begin{gathered}
 \tau_{\text{traj}}=0.3 ,\\
 \Delta \tau =0.004 ,
\label{par}
\end{gathered}\end{equation}
where $\Delta \tau$ is the step size of the molecular dynamic evolution, and
$\tau_{\text{traj}}$ the length of each trajectory.  We performed a rather long
simulation, with a total length after thermalisation of $\tau\simeq 500$ in
units of fictitious time. Fig.~\ref{fig1} shows the Monte Carlo time-history of
the topological charge, as determined after 25 cooling sweeps of a
pseudo-heat-bath algorithm at $\beta=\infty$. These data show clearly that the
HMC is unable to change the global topological modes efficiently, leading to
very long auto-correlations. The value of the topological charge got stuck at
around $Q\simeq -2$, and its value averaged over all configurations generated
so far is decidedly non-zero: $\langle Q \rangle = -1.7(4)$.  A very rough
estimate of the integrated auto-correlation time $T_Q$ from a blocking analysis
of the data gives $T_{Q}\gtrsim 2\times 10^2$ in units of molecular dynamics
time.  The simulations were performed on the 25 Gflops APE tower, with around
$50\%$ efficiency. On this machine $T_{Q} \simeq 200$ corresponds to about
three days of computer time, a considerable amount.  Notice that most
simulations at comparable values of $\beta$ and $m$ presented in the literature
have $\tau \simeq 100$.

To further investigate the behaviour of the HMC, we have performed simulations
with larger quark masses, $m=0.035$ and $m=0.05$, and in the quenched case,
which represents the large quark mass limit of full QCD. Quenched simulations
were performed at $\beta=6.0$, and here HMC has been compared with one of the
best available local algorithms, the over-relaxed heat-bath. One cycle of this
algorithm consists of 5 microcanonical updates followed by a pseudo-heat-bath
update.  We mention that in pure gauge theory a local version of the HMC can be
constructed~\cite{LHMC}, which performs better than the HMC algorithm, but
which cannot be extended to fermions.

\begin{table}[tb]
\begin{small}
\begin{center}
\caption{Results from, and parameters used for the Hybrid Monte Carlo (HMC)
  runs, and, in the last line, the heat-bath over-relaxed (HBOR) run.  The full
  QCD runs with four staggered fermions were performed on a $16^{3}\times 24$
  lattice, the pure SU(3) runs on a $16^{3}\times 16$ lattice. The length of
  each trajectory in fictitious time is $\tau_{\text{traj}}$, the total length
  is $\tau_{\text{total}}$ in units of fictitious time and $\Delta\tau$ is the
  step size.  The last three columns give the integrated auto-correlation time
  of $Q$ in units of fictitious time and wall clock hours on the 25 Gflops
  Quadrics, as well as the auto-correlations for the plaquette ($T_{\Pi}$). 
  For the HBOR algorithm the auto-correlations are given in numbers of cycles
  and wall clock seconds.
}
\label{tab:param}
\vspace{\baselineskip}
\begin{tabular*}{\textwidth}{cc@{\extracolsep{\fill}}cccccccc}
\hline\hline
\multicolumn{1}{c}{$\beta$}&
\multicolumn{1}{c}{$m$}&
\multicolumn{1}{c}{$\Delta \tau$}&
\multicolumn{1}{c}{$\tau_{\text{traj}}$}&
\multicolumn{1}{c}{$\tau_{\text{total}}$}&
\multicolumn{1}{c}{$\langle Q \rangle$}&
\multicolumn{1}{c}{$\chi\times 10^{5}$}&
\multicolumn{1}{c}{$T_{Q}(\tau)$}&
\multicolumn{1}{c}{$T_{Q}({\text{hrs}})$}&
\multicolumn{1}{c}{$T_{\Pi}({\tau})$}\\
\hline\hline
5.35 & 0.010   & 0.004 & 0.3  & 450  & -1.7(4) &          & $\gtrsim$ 200 & $\gtrsim$ 72 & 1.2 \\ 
5.35 & 0.035   & 0.005 & 0.3  & 180  & -0.8(4) & 2.4(9)   & 15   & 2    &2.5  \\ 
5.35 & 0.05    & 0.006 & 0.3  & 300  & -0.3(7) & 11(2)    &  6   & 0.3  &2.2  \\ 
\hline                                                   
6.00 &$\infty$ & 0.01  & 0.3  & 2000 & 0.5(1.3)& 8.1(1.5) &  320 & 0.9  & 1.7 \\ 
6.00 &$\infty$ & 0.01  & 0.6  & 4500 & -0.2(7) & 6.3(1.0) &  140 & 0.4  & 1.7 \\ 
6.00 &$\infty$ & 0.016 & 0.16 & 8960 & -1.0(5) & 6.1(1.9) &  312 & 0.67 & 2.8  \\ 
6.00 &$\infty$ & 0.016 & 0.32 & 3200 & -0.3(7) & 5.8(1.1) &  180 & 0.34 & 1.9 \\ 
6.00 &$\infty$ & 0.016 & 0.64 &23600 &  0.2(2) & 6.1(5)   &   72 & 0.12 & 2.4  \\
6.00 &$\infty$ & 0.016 & 0.96 & 9000 &  0.2(2) & 4.6(6)   &   43 & 0.08 & 4.1(27s)  \\
6.00 &$\infty$ & 0.016 & 1.50 & 9000 &  0.1(2) & 6.9(7)   &   69 & 0.12 & 4.9  \\
6.00 &$\infty$ & 0.016 & 2.00 & 6000 & -0.1(3) & 6.5(5)   &   83 & 0.15 & 5.6  \\
\hline\hline 
6.00 &$\infty$ & \multicolumn{1}{c}{HBOR}  &      &      &  0.04(8)& 5.9(3) &  5.7 & 0.0014 & 0.9(0.9s)  \\ 
\hline\hline 
\vspace{\baselineskip}
\end{tabular*}
\end{center}
\end{small}
\end{table}

In Table~\ref{tab:param} we give the parameters used in our HMC simulations and
the corresponding estimates of the integrated auto-correlation time of the
topological charge $T_{Q}$, given in molecular dynamics time units and in CPU
hours.  $T_{Q}$ has been estimated from a binning procedure, as well as
directly where the statistics were sufficient to be able to calculate the
integrated auto-correlation time. In these cases both agree. The binning
estimates should be regarded as a lower limit for $m=0.01$, and with a possible
uncertainty of 10-20\% for all other values. This is sufficient to our purpose.
The results for the quenched case must be compared with the auto-correlation
time of the local over-relaxed algorithm which is $T_{Q}\simeq 5.7$ cycles, or
$T_{Q}\simeq 5.2$ seconds in CPU time.

There are three major results that have emerged from this work:
\begin{enumerate}
\item In HMC simulations the integrated auto-correlation time of the global
  topological charge  decreases as the length of the trajectory
  increases. If expressed in terms of computer time, however, there is
  a decrease in performance for trajectories longer than one unit in molecular
  dynamics time. The best step size in pure SU(3) seems to be one
  giving an acceptance rate around 70\%.
\item The performance of the HMC is already poor for pure gauge theory, i.e. in
  the limit $m\rightarrow \infty$.  At $\beta=6.0$, it turns
  out to be about two orders of magnitude slower in decorrelating $Q$ than local
  over-relaxed algorithms.  For our `best' set of HMC parameters we find that
  the plaquette decorrelates about 30 times slower than the heat-bath
  over-relaxed algorithm, which agrees with the results of~\cite{gupcomp}.  The
  plaquette auto-correlation time in units of $\tau$ is constant, though.
  
  For our `best' parameters in pure SU(3) $Q$ decorrelates 60 times slower with
  the HMC than with the heat-bath over-relaxed.  Furthermore the
  auto-correlation of $Q$ is far more sensitive to the choice of parameters
  than the plaquette, increasing dramatically in units of $\tau$, and of course
  in real time as well, the more the parameters differ from optimal.
\item The auto-correlation time $T_{Q}$ rapidly increases with decreasing quark
  mass, in terms of both CPU time (which is also due to critical slowing down
  in the inversion of the fermion matrix) and molecular dynamic time. 
  The increase in real time required appears to scale as 
  $T_{Q}=1/m^{\alpha}$ with $\alpha\approx 3 \text{ to } 5$. 
  This is considerably more than the  $\alpha\approx 2.5$ expected from using 
  $1/m^{2}_{\pi}=1/m$ as the relevant physical quantity~\cite{SGacc} causing
  slowing down, 
  a factor $1/m$ from the matrix inversion, and another $\approx 0.5$ from the
  change in step size and acceptance rate. 
  
  The auto-correlation time for the plaquette, on the other hand, remains
  similar in units of $\tau$. So the increase in real time required for the
  topology is considerably more than the increase in real time required for
  smaller quark masses that one has previously learnt to
  expect~\cite{gupcomp,bitcomp} based on studies of more local correlators like
  the plaquette.
  
  In our rather long HMC simulation at $\beta=5.35$ and $m=0.01$ ($\tau\simeq
  500$ in molecular dynamic time unit) the sampling of the topological modes
  appears not to be correct, making the determination of $\chi$ impossible.  We
  cannot even estimate how long the run should have been to get $\chi$.  At
  $m=0.035$ the sampling of $Q$ turned out to be sufficiently long to allow a
  calculation of $\chi$.
\end{enumerate}

In conclusion we have shown that the HMC algorithm has serious problems
decorrelating global topogical modes, more serious than those associated with
commonly studied quantities. For full QCD the algorithm appears to slow down 
as roughly $1/m^{4}$.
We stress again that this warning must be
seriously considered when simulating full QCD.  It is especially important when
studying quantities related to the topology, but may be less relevant in the
calculation of the mass spectrum (with the exception of, for example, the
$\eta^{\prime}$ mass), since an effective decoupling from the topological modes
is expected. Note also that this slowing down is independant of both the number
of fermions and type of fermion simulated, as the underlying dynamics of the
algorithm are the same as the cases studied here.

It may also be possible to improve the performance of the HMC algorithm at
small masses beyond gains coming from tuning the trajectory length and step
size via the use of simulated tempering~\cite{ST}. In this, the quark mass
becomes a dynamical variable in the simulation. This helps in decorrelating the
topological charge both because there are more fluctuations in the charge when
the mass is large, and because the conjugate gradient inversion is faster for
larger masses. This is presently under investigation.  Finally, we mention that
other algorithms, not based on equations of motion, (for example,
L\"uscher's~\cite{luscherferm} multi-boson algorithm), may perform better with
respect to the topological modes than the HMC algorithm. This is also being
examined.

\ack{} 
This project was partially supported by the European Union, contract
CHEX--CT92--0051, and by MURST.  GB was supported by the European Union
{\em  Human Capital and Mobility} program under HCM-Fellowship contract
ERBCHBGCT940665. The authors are particularly grateful to Raffaele Tripiccione
for advice and assistence in using the 512 node APE/QUADRICS in Pisa.



\end{document}